\documentclass[twocolumn,prb,amsmath,amssymb]{revtex4}
\usepackage{graphicx,bm}

\def\<{{\langle}}
\def\>{{\rangle}}
\newcommand{\al}{\alpha}
\newcommand{\be}{\beta}
\newcommand{\de}{\delta}
\newcommand{\D}{\Delta}
\newcommand{\e}{\epsilon}

\newcommand{\w}{\omega}

\newcommand{\s}{\sigma}
\newcommand{\del}{\nabla}

\newcommand{\p}{\partial}

\newcommand{\hatG}{\hat{G}}
\newcommand{\va}{\vec{a}}

\newcommand{\vJ}{\vec{J}}

\newcommand{\mbk}{\mathbf{k}}

\newcommand{\ksn}{{\mbk s n}}

\newcommand{\mbr}{\mathbf{r}}

\newcommand{\mbv}{\mathbf{v}}

\newcommand{\bsphi}[1]{\boldsymbol{\phi}}
\newcommand{\mbf}[1]{\mathbf{#1}}

\newcommand{\bfhat}[1]{\hat{\mathbf{{#1}}}}

\newcommand{\mcal}[1]{\mathcal{#1}}

\newcommand{\til}[1]{\tilde{#1}}

\newcommand{\nn}{\nonumber\\}

\newcommand{\ben}{\begin{equation}}
\newcommand{\een}{\end{equation}}

\begin{document}
\title{Topological transport in a spin-orbit coupled bosonic Mott insulator}
\author{C. H. Wong, R.A. Duine}
\affiliation{Institute for Theoretical Physics, Utrecht University, Leuvenlaan 4, 3584 CE Utrecht, The Netherlands}
\date{\today}
\begin{abstract}
We investigate topological transport in a spin-orbit coupled bosonic Mott insulator.  We show that interactions can lead to anomalous quasi-particle dynamics even when the spin-orbit coupling is abelian.  To illustrate the latter, we consider the spin-orbit coupling realized in the experiment of Lin \textit{et al}. [Nature (London) \textbf{471}, 83 (2011)].  For this spin-orbit coupling, we compute the quasiparticle dispersions and spectral weights, the interaction-induced momentum space Berry curvature, and the momentum space distribution of spin density,  and propose experimental signatures.  Furthermore, we find that in our approximation for the single-particle propagator, the ground state can in principle support an integer Hall conductivity if the sum of the Chern numbers of the hole bands is nonzero. 
\end{abstract}
\maketitle
\textit{Introduction}--Spin-orbit coupled electronic systems that break time reversal symmetry can exhibit the quantum anomalous Hall (QAH) phase, which supports an integer Hall conductivity in the absence of a magnetic field, and was first proposed in a model for graphene.\cite{HaldanePRL88}  At  the single particle level, this is due to the nontrivial momentum space topology of the Hamiltonian, which is characterized by a momentum space Berry curvature that causes in the semiclassical equations of motion an anomalous velocity transverse to external forces, and its integral over the Brillouin zone is a topological invariant known as the Chern number.\cite{HaldanePRL04}  
Since it is now possible to engineer spin-orbit couplings in cold atomic systems,\cite{linNAT11} there have been several recent proposals of cold atomic optical lattice realizations of the QAH phase,\cite{{liuPRA10}} and the related atom topological insulators (TI),\cite{goldmanPRL10} which can be built from two time-reversal conjugate QAH systems.\cite{{hasanRMP10},qiRMP11}All these proposals concern fermions.

While TI's and QAH phases have been extensively studied at the non-interacting single particle level and in the presence of weak interactions, the strongly interacting regime is much less understood.  This regime is relevant for some strongly correlated materials such as the transition metal oxides,\cite{shitadePRL09}  and can be accessed in cold atomic optical lattices,\cite{greinerNAT02} which has the advantage that one can study interaction effects in a perfectly clean environment and create topological phases for bosons, which we will study in this paper. 

So far,  only abelian spin-orbit couplings have been achieved in cold atoms,  as in the experiments of Ref.~[\onlinecite{linNAT11}].   This spin-orbit coupling, equivalent to an equal amount of Rashba and Dresselhaus coupling, has zero Berry curvature in the absence of interactions.   In this paper, we consider the effect of an optical lattice and interactions in the Mott-insulating regime for this system.    We find that the interactions renormalize the momentum space spin texture and generate Berry curvature, resulting in an anomalous velocity for the quasiparticles.

In two dimensional topological insulators such as HgTe quantum wells, one can change the topological phase by tuning a time-reversal breaking gap in the single particle Hamiltonian.\cite{bernevigSCI06}  In a spinful Mott insulator one may consider whether such a gap can arise due to interactions.  We find that this is indeed possible, and that the Chern number can change in the (ferromagnetic) Mott-insulating phase.  If the sum of the hole band Chern numbers is nonzero, we find,  in the approximation that the propagator contains only quasiparticle poles, the ground state can support an integer Hall conductivity. 



\textit{Interacting Berry curvature and Hall conductivity}--
For an interacting system,  the Hall conductivity $\s_H$ can be expressed in terms of the momentum space single particle propagator $G_{\al\beta}(\w,\mbk)$, which reads in two spatial dimensions at zero temperature,\cite{ishikawaNPB87,HaldanePRL04} 
\ben
\s_{H}=-{\e^{ij}\over8\pi^2h}\int d\w d\mbk \, e^{i\w0^+}{\rm tr}[\p_\w \hat{G}\p_i \hat{G}^{-1}\hatG\p_j \hatG^{-1}]\,,
\label{hall}
\een
where here and below $i=(k_x,k_y)$ and $0^+$ denotes a positive infinitesimal, $h$ is the Planck constant (this conductivity gives a mass current as the atoms are neutral).  This expression is a topological invariant in frequency and momentum space\cite{volovikBook03} called the Chern number for an interacting system and is related to the number of edge states, and a nonzero integer value defines QAH phase.\cite{wangPRL10to,qiPRB06} Next we show that this phase can exist for a bosonic Mott insulator that supports quasi-hole excitations with non-trivial Berry curvature, as is typically the case in our approximation.

Deep in the Mott insulating phase,  the relevant excitations are long-lived quasiparticle/quasihole states whose spin-orbit coupling is encoded in the Hermitian matrix structure of $\hat{G}$, which can be expressed the local spin basis as $\hat{G}(\w,\mbk)=\hat{U}\hat{G}_d\hat{U}^\dag$, where $\hat{U}(\w,\mbk)$ is a unitary matrix and $\hat{G}_d$ is a diagonal matrix.  Below we will compute $\hat{G}(\w,\mbk)$ in the Mott insulator.  Using this basis, picking up the contributions from the quasiparticle poles in $\hat{G}_d$ in the $\w$ integration in Eq.~\eqref{hall}, and making the finite temperature generalization of Eq.~\eqref{hall}, the Hall conductivity can be expressed as
$\s_{H}=\sum_{sn}\nu_{sn}/h$, where 
\ben
\nu_{sn}=\int {d\mbk\over2\pi}\,\frac{\til{\mcal{B}}^z_\ksn}{\exp(\e_\ksn/k_BT)-1}\,,
\label{nu}
\een
where $T$ is the  temperature, $k_B$ the boltzmann constant,  $s,n=\pm$ are indices for the spin-orbit and the quasi-particle/hole bands, respectively, and the interacting Berry curvature is given by
\cite{shindouPRL06}
\ben
\til{\mcal{B}}^z_\ksn=\mcal{B}^z_\ksn+\mbf{v}_\ksn\times\bm{\mcal{E}}_\ksn\,,
\label{B}
\een
where $\mbv_\ksn=\p_\mbk\e_\ksn$ are the band velocities, and $\e_\ksn$ are the quasiparticle energies relative to the chemical potential.  The Berry electric field $\bm{\mcal{E}}$ and magnetic field $\mcal{B}^z$, which in $2D$ points in-plane and out-of-plane, respectively, can be expressed in terms of a field strength tensor $(\mcal{B}^z_{s},\mcal{E}^i_{s})=(\mcal{F}^s_{xy},\mcal{F}^s_{\w i})$, where $\mcal{F}^s_{\mu\nu}\equiv\p_\mu\mcal{A}^s_{\nu}-\p_\nu\mcal{A}^s_{\mu}$ with $\mu,\nu=(\w,k_x,k_y)$ is defined in terms the diagonal components of the matrix gauge field $\mcal{A}^s_{\mu}\equiv[i\hat{U}^\dag\p_\mu\hat{U}]_{ss}$. It is convenient to define on-shell matrix rotations and gauge fields in $\mbk$ space by  $\til{U}_\ksn=\hat{U}(\w=\e_\ksn,\mbk)$ and 
\[\til{\mcal{A}}_{\ksn}\equiv i[\til{U}^\dag\p_\mbk\til{U}]_{ss}=[\mcal{A}^s_{\w}\mbv_\ksn +{\mcal{A}}^s_{\mbk}]|_{\w=\e_\ksn}\,,\] then with analogous definitions as above,  $\til{ \mcal{B}}^z_\ksn=\til{\mcal{F}}^s_{xy}$.

In the zero temperature limit, only the quasihole contributions in Eq.~\eqref{nu} with $\e_\ksn<0$  remain,
\ben
\s_H={1\over h}\sum_{s}-{1\over2\pi }\int_{\rm BZ} {d\mbk}\,\til{\mcal{B}}^z_{\mbk s-}\,.
\label{hall1}
\een
where the integral over the Brillouin zone (BZ) is the Chern number, which is a topological property of each band.  If the sum of these numbers is nonzero,  this results in a ground state, integer Hall conductivity.   Thus, although the QAH phase cannot occur for non-interacting bosons, it can occur in the bosonic Mott insulator, or more generally, for any bosonic system with hole bands of which the sum of Chern numbers is nonzero.

\textit{Spin-orbit coupled Bose Hubbard model}--
In the remainder of this paper we compute the Green's function in the Mott insulator and we specifically consider a two dimensional (2D) square optical lattice of bosons with two spin components deep in the Mott-insulating phase at a commensurate filling, with spin-dependent hopping amplitudes. We will describe the system by a  pseudo-spin $1/2$  Bose Hubbard model with onsite repulsive interactions, and write the Hamiltonian as $H=H_0+V$,  where
\begin{align}
H_0&={1\over 2} \sum_{i,\al\beta} U_{\al\beta}a^\dag_{i\al}a^\dag_{i\beta} a_{i\beta}a_{i\al}\,;\nn
{V}&=\sum_{<i,j>} \va_{i}^\dag\hat{t}_{ij}\va_{j}=\sum_\mbk \va_{\mbk}^\dag \hat{h}_\mbk\va_{\mbk}\,,
\end{align}
where  $i$ denotes lattices sites, $<>$ denotes nearest neighbors, $\al,\beta=\pm$ are spin indices, hats denote matrices in spin space and $\vec{a}_\mbk$ and $\vec{a}_i$ denote two-component spinor field operators in momentum and real space, respectively, and $\hat{t}_{ij}$ are hopping matrices.   It is convenient to write $\hat{h}_\mbk=h_\mbk+\mbf{h}_\mbk\cdot\hat{\bm\s}$, where $\hat{\bm{\s}}$ is the vector of Pauli matrices, and $\mbf{h}_\mbk$ is the non-interacting spin-orbit field.  The dispersions of the non-interacting particles are given by $\e_{1\mbk\al}=h_\mbk+\al|\mbf{h}_\mbk|$. We will first derive results for a general $\mbf{h}_\mbk$, and then apply them to the spin-orbit coupling of Ref.~[\onlinecite{linNAT11}].
\begin{figure}[t]
\begin{center}
\begin{tabular}{cc}
\includegraphics[width=.5\linewidth]{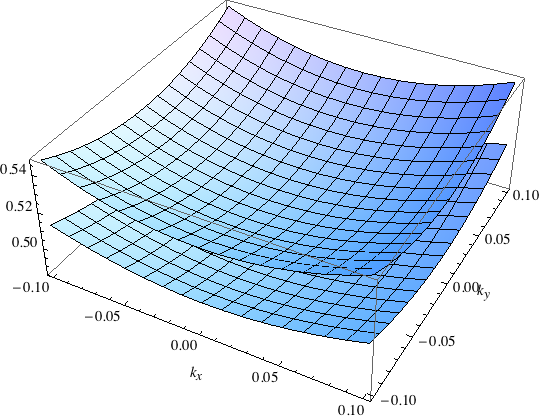}
&\includegraphics[width=.5\linewidth]{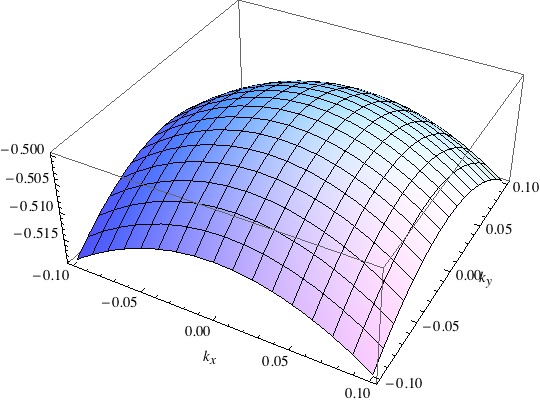}\\
(a)&(b)\\
\includegraphics[width=.5\linewidth]{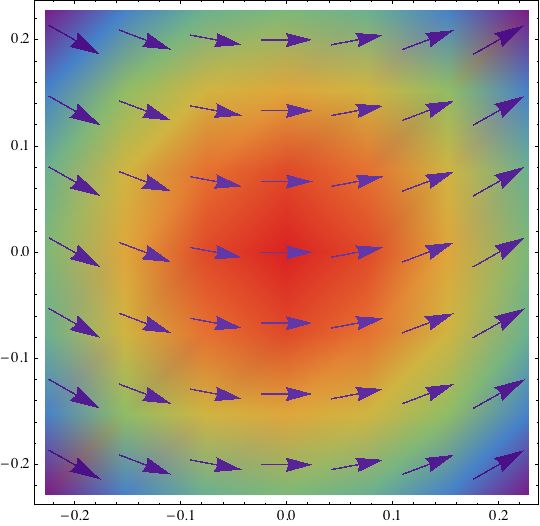}
&\includegraphics[width=.5\linewidth]{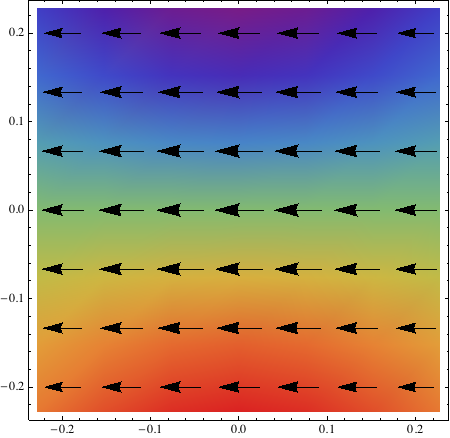}\\
\includegraphics[width=.5\linewidth]{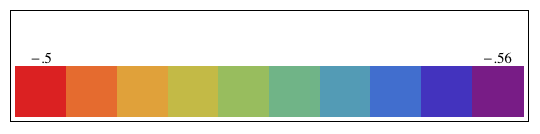}
&\includegraphics[width=.5\linewidth]{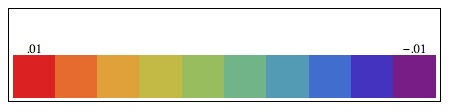}\\
(c)&(d)
\label{spin}
\end{tabular}
\caption{For the filling fraction $(1,0)$: Dispersions with energy in units of $U$ with $\eta=\D=.01$ for 
(a) spin up and down quasiparticles and (b) spin up quasiholes. 
(c) Ground state momentum distribution of spin density $\mbf{s}_\mbk$ for $\eta\sqrt{m/U}=\D/U=.04$.  The vector field represent the in-plane components and the colored density plot represent the z-component.
(d) The ground state Berry electric field $\mcal{E}^x$  and interacting Berry curvature $\til{\mcal{B}}^z$ of the spin up quasiparticle band, plotted as function of dimensionless wave vectors in the region $|k'_{x,y}|<0.2$.}
\label{fig}
\end{center}
\end{figure}


\textit{Strong coupling perturbation theory}--We will compute the single particle green function in the strong coupling limit $V/U_{\al\be}\ll1$, using a perturbation theory in which the hopping Hamiltonian $V$ is the perturbation to the interaction Hamiltonian $H_0$. \cite{fisherPRB89,oostenPRA01}  To this end, consider the grand canonical partition function with external sources $\vJ$, which has the path integral representation (here and below we set $\hbar=1$)
\begin{align}
&Z[\vJ,\vJ^*]=\int \mcal{D}\va \mcal{D}\va^*\,  \exp\left[-S_0[\va]\right.\nn
&\left.+\int d\tau\left(\sum_{<ij>}-\va_i^*\cdot\hat{t}\cdot\va_j+\sum_i\vJ_i^*\cdot\va_i+\va_i^*\cdot\vJ_i\right)\right]
\end{align}
where $S_0=\int d\tau\,\sum_{i\al} [a_{i\al}^*(\p_\tau-\mu_\al)a_{i\al}+H_0(\va)]$ and $\tau$ denotes imaginary time.  We will use it as a generating functional for the correlation functions  given by  $\<a_{i\al} a_{j\beta}^\dag\>=(1/ Z)({\de^2Z}/{\de J_{i\al} \de J_{j\beta}^*})|_{J=0}$.  

We will specify the unperturbed eigenstates in the occupation number basis by the number of spin up and spin down particles per site, $(N_+,N_-)$ with energies per site given by 
\ben
E_{N_+ N_-}=\sum_\al\left[{U_{\al}\over 2}N_\al(N_\al-1)-\mu_\al N_\al\right]+U_{+-}N_+N_-\,.
\label{onsiteE}
\een
The unperturbed imaginary--time ordered correlation function is defined by $\de_{ij}\de_{\al\beta}g_\beta(\tau,\tau')=-\<N_+N_-|\mcal{T}a_{i\al}(\tau')a^\dag_{j\beta}(\tau)|N_+N_-\>$, which is local in space and diagonal in spin, and its Fourier transform reads   
 \begin{align}
g_\al(i\omega_n)
&=\frac{1+N_\al}{i\omega_n-\xi_\al(N_+,N_{-})}-\frac{N_\al}{i\omega_n{-\xi_\al(N_\al-1,N_{-\al})}}\,.
\label{onsiteg1}
\end{align}
where $\xi_\al(N_+,N_-)\equiv E_{N_\al+1,N_{-\al}}-E_{N_\al,N_{-\al}}$ are the excitation energies, and $\w_n$ are bosonic matsubara frequencies.  We will only be interested in the zero temperature limit, and the dependence on $N_\al$ above has been calculated in this limit.    We also define the total particle $g_0$ and spin $g_z$ components by $g_{0}=(g_{+}+g_{-})/2\,,g_{z}=(g_{+}-g_{-})/2$.

Carrying out this perturbation theory,  we find
\ben
-\hat{G}^{-1}(i\omega_n,\mbk;N_\al)=-\hat{g}^{-1}(i\w_n;N_\al)+\hat{h}(\mbk)\,.
\label{Ga1}
\een
Although this expression is first order in hopping, its inverse, the propagator, contains all orders in hopping.

In the following, we will denote by the replacement $i\w_n\to\w$ the analytic continuation of the green functions, which will be necessary to compute the spectral functions and quasiparticle dispersions, and to define the spin texture in $(\w,\mbk)$ space.  Defining components by $-\hat{G}^{-1}=d+\mbf{d}\cdot\hat{\bm{\s}}$, the  spin textures are given by
\ben
\mbf{d}(\w,\mbk;N_\al)=-[g^{-1}(\w;N_\al)]_z\mbf{\hat{z}}+\mbf{h}(\mbk)\,,
\label{da}
\een
and $d(\w,\mbk;N_\al)=-[g^{-1}(\w;N_\al)]_0+h(\mbk)$.   The in-plane texture is unchanged, while the $z$ component is shifted by the $\s_z$ component of the inverse propagator. 

The quasiparticle dispersions are computed by solving $-({G}_d^{-1})_{ss}(\w=\e_{\ksn})=d+s|\mbf{d}|=0$.  The Berry electromagnetic fields are given by
\ben
(\mcal{B}^z_\ksn,\mcal{E}^i_{\ksn})={s\over2}\mbf{\hat{d}}\cdot(\p_x\mbf{\hat{d}}\times\p_y\mbf{\hat{d}},\p_\w\mbf{\hat{d}}\times\p_i\mbf{\hat{d}} ) |_{\w=\e_\ksn}\,.
\label{EB}
\een
where $\mbf{\hat{d}}\equiv\mbf{d}/|\mbf{d}|$ and the expressions above are evaluated at ${\w=\e_\ksn}$.  Defining the quasiparticle spin-orbit field in $\mbk$ space  by $\mbf{f}_\ksn=\mbf{d}(\w=\e_\ksn,\mbk)$,  it is readily verified that  $\til{\mcal{B}}^z_\ksn/2\pi=(s/4\pi){\bfhat{f}_\ksn\cdot\p_x\bfhat{f}_\ksn\times\p_y\bfhat{f}}_\ksn$, and thus the Chern number is the integer winding number of $\mbf{f}_\ksn$.

From the simple poles of $\hat{G}=(d-\mbf{d}\cdot\hat{\bm{\s}})/({d^{2}-\mbf{d}^{2}})$, we find that quasiparticle peaks in the spectral function $\hat{A}(\w)=i(\hat{G}(\w+i0^+)-\hat{G}(\w- i0^+))$ are given by,
\ben
\hat{A}(\omega,\mbk)=-2\pi\sum_{s n}
Z_\ksn{1+s\bfhat{f}_\ksn\cdot\mbf{\hat{\bm{\s}}}\over2}\de(\w-\w_\ksn)\,,
\een
where $Z_{\ksn}=[\p_\w(d+s|\mbf{d}|)|_{\omega=\e_{\ksn}}]^{-1}$ are the quasiparticle weights.  The numerator above is a spin projection operator that can be written as ${\chi_\ksn\chi_{\ksn}^\dag}$, where $\chi_\ksn$ is the normalized spinor part of the quasiparticle wave functions which points in the direction $\bfhat{f}_\ksn$.  We have checked that the trace of the spectral function contains only these quasiparticle peaks.  The momentum distributions of particle and spin density are given by $\{n_\mbk,\mbf{s}_\mbk\}=\int(d\w/2\pi) {\rm tr}[\{1,\hat{\bm{\s}}\}\hat{A}(\w,\mbk)]/( e^{\w/k_BT}-1)$, which in the ground state (cf. Eq.~\eqref{hall1})  contains only quasihole contributions given by $\{n_\mbk,\mbf{s}_\mbk\}=\sum_s Z_{\mbk s-}\{1,s\bfhat{f}_{\mbk s-}\}$.

Next, we consider when the interaction induced spin-orbit texture $\bfhat{f}_\ksn$ can have a nonzero winding number.  For this to be true, $\mbf{h}_\mbk$ must have a nonzero in-plane winding number and hence contains vortices.  We denote the position of these vortices by $\mbk_i$, and the gaps at the vortex cores which determine the Chern number are given by
\ben
 {f}^z_{sn}(\mbk_i)=(g^{-1})_z(\e_{sn}(\mbk_i))+h_z(\mbk_i)\,.
\label{fz}
\een
To have a nonzero winding number, the gaps need to have different signs at different $\mbk_i$'s.   Naively, since $(g^{-1})_z$ is of order $U_{\al\beta}$, one might expect it to be the dominant term and set a large uniform gap.  However, in our perturbation theory, the quasiparticle energies are always close to the onsite particle/hole spectrum which are the zeros of $\hat{g}^{-1}(\w)$, so that as long as the the onsite spectrum is nearly spin-symmetric, \footnote{This requires that $N_\al\neq0$,  so that there are quasihole for both spins.} $(g^{-1})_z(\e_{sn}(\mbk_i))\sim O(\hat{h}_1)$.  Therefore, the winding number is determined by the competition between the terms in Eq.~\eqref{fz}, and may differ than the non-interacting one determined by $h_z(\mbk_i)$.  



%
%

\textit{A simple case}--To illustrate the relevant physics in the simplest setting, we turn off the inter-spin interaction, setting $U_{+-}=0$ and $U_{++}=U_{--}=U$. 
We use $U$ as the unit of energy and define the dimensionless quantities $\mbk'=\mbk/\sqrt{2mU}$. The excitation energies are $\xi_\al(N_\al)=UN_\al-\mu_\al$ and the zeroth order ground state has filling factors $N_\al$ when $N_\al-1<{\mu_\al/ U}<N_\al$.  Setting ${\mu_\al}=U(N_\al-1/2)$,\footnote{The chemical potentials 
and will generally acquire perturbative $O(V/U)$ corrections.} 
the inverse total particle and spin propagator is
\begin{align}
-\{(g^{-1})_0,{(g^{-1})_z}\}&=\frac{U  (4 \omega ^2-1)}{2 (1+2 N_++2 \omega ) (1+2 N_-+2 \omega )}\nn
&\{1+N_++N_-+2 \omega,N_--N_+) \}\,.
\label{invg}
\end{align}

In the cases of equal filling fractions $(N_+,N_-)=(N,N)$,  $(g^{-1})_z=0$ so that there are no corrections to the spin texture.  The simplest filling fraction which gives an out-of-plane spin texture is $(1,0)$.  This filling represents a ferromagnetic ground state, which is generally present in a bosonic Mott insulator.  The quasiparticle dispersions can be expressed implicitly as 
\ben
{\e_\ksn\over U}=\frac{1}{4} \left(-1+2 \til{\e}_\ksn+n\sqrt{9+20 \til{\e}_\ksn +4 \til{\e}_\ksn ^2}\right)\,,
\label{disp}
\een
where $\til{\e}_\ksn={(h_\mbk+s|\mbf{f}_\ksn|)/ U}$, $s,n=\pm1$.  Note that there is no spin down quasihole band $(s,n)=(-,-)$ because there are no spin down atoms in the unperturbed $(1,0)$ ground state.  Thus, the ground state momentum distribution of spin density is given by $\mbf{s}_\mbk=Z_{\mbk +-}\bfhat{f}_{\mbk +-}$. 

To give an example of hole bands with nonzero Chern number, we consider one component of the BHZ Hamiltonian which describes HgTe quantum wells,\cite{bernevigSCI06} and we take parameters such that  $h=2t(\cos{k_x}+ \cos{k_y})$, $\mbf{h}=2t_{\rm so}(\sin{k_x},-\sin{k_y},0)$.  In the absence of interactions these parameters give zero Chern number.   However, by computing the gaps in Eq.~\eqref{fz}, we find for the filling factor $(1,2)$ with $t/U\sim.01$, and an additional onsite spin splitting of $\de/U\sim .01$ that the Chern numbers are $\pm1$, so the sum of the Chern numbers is zero for this model.  


%

\textit{An abelian SO coupling}--We now consider the spin-orbit coupling of Ref.~[\onlinecite{linNAT11}] by taking a Hamiltonian that at low energy is specified by   $h_\mbk=\mbk^2/2m$ and $\mbf{h}_\mbk=(\D_x,\eta{k_x},0)$, where $m$ is the effective mass,\footnote{It has order of magnitude $m\sim1/|t|a_0^2$, where $a_0$ the lattice spacing, $|t|$ is the order of magnitude of the hopping matrix} $\eta$ is the spin-orbit coupling and $\D_x$ is an applied detuning.     We plot the quasiparticle dispersions in Fig.~\ref{fig}(a,b) and $\mbf{s}_\mbk$ in Fig. \ref{fig}(c).  The in-plane components of $\mbf{s}_\mbk$ are qualitatively the same as $\mbf{h}_\mbk$, but interactions generate an out-of-plane component.  Although the non-interacting Berry curvature vanishes, $\mbf{\hat{h}}\cdot\p_x\mbf{\hat{h}}\times\p_y\mbf{\hat{h}}=0$, the interacting Berry electric field and curvature $\mcal{E}^x_\mbk$ and $\til{\mcal{B}}^z_\mbk=-v^y_{\mbk}\mcal{E}^x_{\mbk }$, respectively, are nonzero, which we plot in Fig.~\ref{fig}(d).  The Chern number is still zero because $\mbf{h}$ has zero in-plane winding, and this is verified in our computation because $\til{\mcal{B}}^z_\mbk$ is antisymmetric under inversion and hence vanishes upon integration.   There are still nontrivial physical consequences due to the quasiparticle anomalous velocity $\dot{\mbr}_{\ksn}=\til{\mcal{B}}^z_\ksn(-\del\phi\times\mbf{\hat{z}})$ which gives an edge current in the presence of a trapping potential $\phi$. Furthermore, it was shown in Ref.~[\onlinecite{shindouPRL06}] for the Fermi liquid that the quasiparticle weight is renormalized by a factor $Z'\sim-\bm{\mcal{E}}\cdot\del\phi$. Since the only key assumption was a hermitian self energy matrix, which is true for our case, we expect this to be valid.  



\textit{Experimental signatures}--There are many experimental signatures of the interaction induced Berry curvature and topological transport that we studied in this paper. The ground state particle and spin density can be measured using RF spectroscopy.\cite{feldNATL11}  and time of flight measurements.\cite{albaPRL11}    The quasiparticle dispersions and spectral weights enter into the density--density response function which can be measured with Bragg spectroscopy.\cite{oostenthesis}    A nonzero anomalous Hall conductivity causes transverse oscillations in collective modes which can be seen by absorption imaging of the density.\cite{bijlPRL11}  Furthermore,  bulk and edge quasiparticle transport can be probed directly by exciting particle-hole  with lasers.



\textit{Conclusion and outlook}--We have shown that in a spinful bosonic Mott insulator, interactions significantly modify the quasiparticles spin-orbit momentum-space texture.  In our strong-coupling perturbative approximation,  we have computed the interacting Green function for a general spin-orbit coupled hopping Hamiltonian, and show that the ground state can in principle support an integer Hall conductivity given by the sum of the hole Chern numbers.  For the optical lattice version of the experiment in Ref.~[\onlinecite{linNAT11}], we computed quasiparticle dispersions, Berry curvature, and momentum distributions of spin density.  We hope this will motivate experiments to measure these quantities, and to create the spin-orbit couplings with in-plane vortex textures which we have predicted will exhibit topological transport.

It seems to be a generic property of two band models, which have opposite signs of the Berry charge, that their Chern numbers tend to cancel.   For future work, we will investigate whether it is possible to find a model with a nonzero sum of hole Chern numbers, for example, in a three-band model.  According to the arguments presented in Ref.~[\onlinecite{senthilCM12}], the Hall conductivity for bosons should be quantized in even integers.  Further research is needed to investigate the mutual consistency of these arguments with our approximation and the zero temperature Hall conductivity that follows from it.   



By choosing the $(1,0)$ filling fraction, we have in a sense broken time reversal symmetry ``by hand.''  In the presence of inter-spin interactions $U_{+-}$, time reversal symmetry will be broken spontaneously because the ground state will have magnetic order.\cite{colePRL12}  This can readily be included in our theory, where Eq.~\eqref{Ga1} still holds but $\hat{g}^{-1}$ acquires off-diagonal components.\cite{grassPRA11}  A uniform ferromagnetic state gives qualitatively similar results to the case studied in this paper.  Furthermore, even in the presence of magnetization textures, our momentum-dependent results will  be approximately valid for momenta larger than the inverse of the typical length scale of the texture.  In general, textures will result in interplay between momentum space and real space berry curvature, which motivates further study.


Finally, in this work we have considered only the Mott-insulating phase.  We leave the analysis of the Mott insulator--superfluid transition to future work.  We also intend to investigate the bulk and edge state transport properties of these systems in more detail. 


We thank Henk Stoof, Allan MacDonald, Yaroslav Tserkovnyak, and Ari Turner for valuable discussions.  This work was supported by Stichting voor Fundamenteel Onderzoek der Materie (FOM), the Netherlands Organization for Scientific
Research (NWO), by the European Research Council (ERC) under the Seventh Framework Program (FP7).


\end{document}